\documentclass[preprint,12pt]{elsarticle}
\usepackage{graphicx}
\usepackage{amssymb} 
\usepackage{amsmath}    %needed for \begin{cases}...\end{cases}
\usepackage{latexsym}   %needed for \Box
\usepackage{bm}
\usepackage{times}

\begin{document}
%\tableofcontents
\begin{frontmatter}

  \title{Incompleteness theorem for physics}

  \author[jm]{John M. Myers\corref{cor1}}
\ead{myers@seas.harvard.edu}
\author[hm]{F. Hadi Madjid}%% \address[label2]{<address>}
\ead{gailmadjid@comcast.net}
\cortext[cor1]{Corresponding author}
\address[jm]{Harvard School of Engineering and Applied  
Sciences, Cambridge, MA 02138, USA}
\address[hm]{82 Powers Road, Concord, MA 01742, USA}

%\date{Draft of \textbf{\today;} \textbf{03-20prl.tex}}% It is always \today, today,

\begin{abstract} %For prl, 600 character max, spaces included.
 We show how G\"odel's incompleteness theorems have an analog in quantum
theory.  G\"odel's theorems imply endless opportunities for appending
axioms to arithmetic, implicitly showing a role for an entity that writes
axioms as logically undetermined strings of symbols.  There is an analog of
these theorems in physics, to do with the set of explanations of given
evidence.  We prove that the set of explanations of given evidence is
uncountably infinite, thereby showing how contact between theory and
experiment depends on activity beyond computation and measurement---a
physical activity of logically undetermined symbol handling.
\end{abstract}


\end{frontmatter}
\maketitle
\section{Introduction}
People writing and reading symbols act as symbol handlers, for example
in making use of numerals and symbols of an alphabet.  In mathematical logic,
symbol handling has attracted serious attention, notably with
G\"odel's incompleteness theorems.  By analyzing the system of symbols
involved in the most elementary branch of mathematics, the arithmetic of
natural numbers 0, 1, 2, \ldots, G\"odel proved that there are always
questions in mathematical logic, the answers to which are logically
undetermined, calling for additional axioms which in turn generate more such
questions \cite{godel}.  Then came Turing's formulation of the concept of
computability in terms of symbol handling, leading to his demonstration of
the prevalence of functions of natural numbers that are {\it uncomputable}.
G\"odel and Turing's attention to the manipulation of symbols in mathematics
opened a field of inquiry into logically undetermined aspects of mathematics,
with implications for other fields of endeavor that make use of mathematics,
including physics.

In physics, strings of symbols express both evidence and the formulas by
which that evidence is explained, formulas that lead to predictions.
Recently came the recognition that the expression in symbols has interesting
implications for physics \cite{aop05,aop16}.  In earlier work, we proved that
quantum-theoretic explanations of given evidence are logically undetermined
by the evidence, leading to the necessity of guesswork in arriving at an
explanation.  In the next section we prove a stronger proposition: the set of
explanations that fit given evidence is uncountably infinite. Implications of
the proposition follow.  As discussed in Sec.\ \ref{sec:3}, a corollary to
the proof implies an endless open cycle in which symbol-handling agents guess explanations
and test their implications.  Sec. \ \ref{sec:4} discusses the notion of a
symbol handler as exercising capabilities to compute and to guess. To distinguish
between the ultimate reach of computation and other capabilities, we
introduce a theoretical symbol handler that has the full computational
capacity of a Turing machine.  In the context of a Turing machine, as
discussed in Sec.\ \ref{sec:5}, the uncountable set of explanations has
strong implications in relation to what can and cannot be computed.
Sec.~\ref{sec:6} views agency as pervading the material world.  In
Sec.~\ref{sec:7} we make a few remarks toward future directions for
investigation.

\section{Uncountable set of explanations of given evidence}\label{sec:2}
As represented in quantum theory, an experiment involves many trials,
with an outcome occurring for each trial.  Eschewing saying what outcome will
occur at a particular trial, quantum theory speaks only of probabilities of
outcomes.  In explaining probabilities of outcomes, one views a trial as
consisting of the preparation of a state, e.g.  expressed by a density
operator $\rho$, and a measurement, expressed by a Positive-Operator-Valued
Measure (POVM), here denoted $M$.  For some outcome space $\Omega$, with
$\tilde{\Omega}$ a $\sigma$-algebra of subsets of $\Omega$, one then has for
each outcome $\omega \in \tilde{\Omega}$ a non-negative operator $M(\omega)$,
the measurement operator for outcome $\omega$.  Both the density
operator and the measurement operator are operators on some (finite- or
infinite-dimensional) Hilbert space $\mathcal{H}$.  By the Born trace rule,
the explanation $(\rho,M)$ implies a probability for the outcome $\omega$:
\begin{equation}\label{eq:one}
\Pr(\omega)=\mathrm{tr}[\rho M(\omega)].  
\end{equation}
The whole outcome space $\Omega$ is the union of all outcomes. By definition, $\Omega$ is an element of the $\sigma$-algebra $\tilde{\Omega}$, and $M(\Omega) = \bm{1}$, the identity operator on $\mathcal{H}$.

Here is the issue.  Books on quantum mechanics teach us to calculate
probabilities from given states and measurement operators, using the Born
trace rule; however, experiments with unexpected results present the
``inverse situation'' (as in quantum decision theory).  One is given
probabilities abstracted from evidence on the workbench, and one seeks
``blackboard'' explanations in terms of states and measurement operators.  In
practice, an investigator's education and habits of thought may limit the
choice of explanations, but logic alone leaves open a vast region of
explanations that exactly fit the given probabilities.
\begin{quote}
\textbf{Theorem}:
The set of inequivalent explanations that exactly fit given probabilities is
uncountably infinite.
\end{quote}

\noindent{\it Proof:} The idea behind the proof is to exploit tensor products
of Hilbert spaces.  For example, an experiment with two detectors can be
explained using measurement operators that are tensor products, one factor of
the tensor product for each detector.  Ignoring outcomes of one detector
coarsens the experiment; one explains the remaining evidence by setting the
measurement operator for the ignored factor to the identity operator.  In
this way a explanation of evidence is condensed into a simpler explanation of
a condensation of the evidence.  This process can be reversed: any
explanation of evidence from an experiment can be seen as a condensation of
any of a multitude of possible extensions of the experiment.  Such an
extension entails augmenting the Hilbert space $\mathcal{H}^{(0)}$ by another
Hilbert-space factor $\mathcal{H}^{(1)}$ to get
$\mathcal{H}^{(0)}\otimes\mathcal{H}^{(1)}$.

The proof proceeds in two steps. In the first step, we prove that for any
explanation there are always two more inequivalent explanations, as follows.
Suppose that $\tilde{\Omega}^{(0)}$ is a $\sigma$-algebra of outcomes for an
explanation of an experiment, and that probabilities $\Pr(\omega^{(0)})$ are
abstracted from results for each outcome $\omega^{(0)} \in
\tilde{\Omega}^{(0)}$.  Suppose further that one is given an explanation
$(\rho^{(0)},M^{(0)})$ with the operators on a Hilbert space
$\mathcal{H}^{(0)}$ that exactly imply the given probabilities.  Then there
are always two more inequivalent explanations that give the same
probabilities.  Both of these explanations involve a tensor product Hilbert
space of $\mathcal{H}^{(0)}$ with a second factor $\mathcal{H}^{(1)}$.  The
explanations for outcome $\omega^{(0)}$ are
$\left(\rho^{(0)}\otimes\rho_1^{(1)},M^{(0)}(\omega^{(0)}) \otimes
M^{(1)}(\Omega^{(1)}\right)$ and
$\left(\rho^{(0)}\otimes\rho_2^{(1)},M^{(0)}(\omega^{(0)}) \otimes
M^{(1)}(\Omega^{(1)})\right)$, where $\rho_1^{(1)}$ and $\rho_2^{(1)}$ are
distinct density operators on $\mathcal{H}^{(1)}$.  Because of the equality
$M^{(1)}(\Omega^{(1)})=\bm{1}^{(1)}$, these explanations are blind to the
second factor of the density operator, which is why they give the same
probabilities as does the given explanation.  But the two explanations are
inequivalent, because they extend to an expanded experiment which provides
for attending to two or more distinct outcomes in $\tilde{\Omega}^{(1)}$, for
which the two extended explanations imply different probabilities.

Now for the second step.  As a mathematical construct, the augmentation by one more tensor-product factor can be repeated without end,
resulting in a set of mutually inequivalent explanations, one for each of the infinite
sequences $j_1,j_2,\ldots$:
\begin{equation} \label{eq:two}
  \left(\rho^{(0)}\otimes \bigotimes_{n=1}^\infty \rho^{(n)}_{j_n}, M^{(0)}(\omega^{(0)})
  \otimes \bigotimes_{n=1}^\infty
  M^{(n)} (\Omega^{(n)})\right)
\end{equation}
The explanations with the infinite sequence of factors differ independently
for each factor.  Thus the set of explanations has the cardinality of the set
of infinite binary fractions, a set that by the Cantor diagonal argument is
uncountable, meaning it cannot be mapped 1-to-1 to the set of natural
numbers, or, less formally, that it cannot be listed even on an infinitely
extendable Turing tape. Expression (\ref{eq:two}) displays an uncountable
subset of the set of inequivalent explanations that match a given
evidence. It follows that the whole set of inequivalent explanations must also
be uncountable.  Q.E.D.

\begin{quote}
  {\bf Corollary}: There is no logical ground to exclude any of the
  uncountable set of potential explanations of given evidence prior to
  additional evidence not yet on hand.
\end{quote}

\noindent\textbf{Remarks}:
\begin{enumerate}
\item  The multiplicity of explanations has nothing to do with
imperfections in the fit between evidence and its explanation; as proved
above, it holds even in the ideal case in which one demands an exact
fit. Requiring only an approximate fit, as is common practice, makes room for
even more explanations of any given evidence.
\item Quantum-state tomography claims to determine a quantum state from
  evidence.  To this claim we respond that quantum state tomography assumes
  that the measurement operators are known, and, by the Theorem, this knowledge
  cannot be obtained from evidence alone.  In addition, quantum state
  tomography assumes some finite dimension of the Hilbert space, a dimension
  underivable from evidence.
\item It can also be noted that evidence is expressed by probabilities that
  are functions of parameters; we think of parameter values as settings of
  knobs, such as a knob by which to vary a magnetic field strength or a knob
  to vary the angle of a polarizer.  Then not only the evidence but also the
  density operator and the POVM depend on knob settings.  The proof of the
  Theorem goes through the same way for each knob
  setting\cite{aop05,tyler07}.
\item For the physicist struggling to come up with {\it some} explanation of
  unexpected evidence, worrying about the possibility of other explanations
  may seem superfluous; yet being aware that something outside of
  logic is required to make an explanation liberates one from futile efforts
  to derive an explanation by logic alone.
\end{enumerate}

After introducing Turing machines, we will discuss implications of the
uncountability of the set of explanations of given evidence in more detail,
but here is a first hint.  An impediment to assigning a suitable probability
measure to the set arises, because the set of explanations is both
uncountable and devoid of a natural metric.  The real numbers are uncountable
but come with a metric, and the metric allows for probability measures that
in effect assign a probability not to a particular real number but to an
interval of real numbers.  (Intervals expressing values of measurable
variables are appropriate because two real numbers that are sufficiently
close are experimentally indistinguishable.)  A countably infinite set
requires no metric in order for non-zero probabilities to be assigned to its
elements, because its elements can be ordered by natural numbers.  For any
$0<r<1$, an symbol-handling agent can assign a probability $(1-r)r^n$ to the
$n$-th element of the countable set.  (As probabilities must, these sum to
$1$.

In contrast to the metric structure of real numbers and the countability of
natural numbers, explanations of given evidence are discrete and uncountable.
They are like infinite sequences of distinct digits.  If a digit in one
sequence differs from a corresponding digit in another sequence, the
sequences are distinct, regardless of how far along the sequence the
difference appears.  Because of its discrete topology on an uncountable set,
we have the following.
\begin{quote}
  \textbf{Proposition 1}: Any probability measure on the set of explanations
  of given evidence must assign zero probability to all but a countable subset
  of its elements.
\end{quote}
By Prop.\ 1 and and the Corollary to the Theorem, we have
\begin{quote}
  \textbf{Proposition 2}: Neither an explanation nor a probability measure on explanations can be be logically determined by the evidence explained.
\end{quote}

\section{Open cycle of guessing and testing}\label{sec:3}
Out of a potential for uncountable inequivalent explanations, physicists
write down particular explanations.  By the Corollary, the writing of symbols
that introduce an explanation of given evidence takes something beyond
logic and evidence.  This `something' can reasonably be called a {\bf
  guess}. The guess, neither derived mathematically on the blackboard nor
generated from an experiment on the work bench, comes from ``somewhere
else.''   Guessed explanations, expressed in mathematical symbols, feed into
the development of experimental devices and into the design of future
experiments.  Different guessed explanations lead to different experimental
designs, leading to different bodies of evidence that call for more
explanations and hence more guesses.  Any particular explanation is
essentially certain to require revision when tested over enough of its
extensions.  We thus arrive at
\begin{quote}
  \textbf{Proposition 3}: The regularities that physicists find in the
  material world defy any final expression and any final explanation.
\end{quote}
With Prop.~3, we see that quantum theory implies an endless evolution of
physics as an activity of symbol-handling agents ``guessing and testing'' in
an open cycle with no possibility of completion.

Attending to symbol handling clarifies a distinction between an occurrence of
an event of low probability and what we term a {\it surprise}. Via the Born
trace rule, an explanation assigns probabilities to a (possibly infinite)
list of possible outcomes. One can think of this list of possibilities as a
string of symbols recorded in some agent's computer file.  In this way we
link the concept of {\it possible} to the contents of an agent's memory.  In
contrast, our dictionary defines {\it possible} as: able to happen although
not certain to \ldots \cite{OxDict}, without reference to an agent or to any
memory.  By referencing possibilities to an agent's list we define a {\it
  surprise} as a reaction of an agent to an event not on the list of
possibilities---the occurrence of ``an unknown unknown.''  Historically, (as
in the disaster at the Three-Mile Island nuclear plant), people experience
outcomes not on their lists of possibilities, and they react to such
outcomes, thereby experiencing surprise. (Experiencing surprise is distinct
from facing measurement uncertainty.  {\it Uncertainty}, denotes a spread in
a probability distribution ensuing from some measurement model \cite{vim},
involving a list of possible outcomes, and not the introduction of a
possibility not in the list.  Symbol-handling agents operating in the open
cycle of guessing and testing encounter surprises that call for revising
their guesses and the explanations that ensue from them; their assumptions
evolve. In 2005 we wrote of ``the tree of assumptions''\cite{aop05}:
\begin{quote}
Some guesses get tested (one speaks of {\em hypotheses}), but testing a guess
requires other guesses not tested.  By way of example, to guide the choice of
a density operator by which to model the light emitted by a laser, one sets
up the laser, filters, and a detector on a bench to produce experimental
outcomes.  But to arrive at any but the coarsest properties of a density
operator one needs, in addition to these outcomes, a model of the detector,
and concerning this model, there must always be room for doubt; we can try to
characterize the detector better, but for that we have to assume a model for
one or more sources of light.  When we link bench and blackboard, we work in
the high branches of a tree of assumptions, holding on by metaphors, where we
can let go of one assumption only by taking hold of others.
\end{quote}

\section{Agency in physics}\label{sec:4}
With the recognition of an open cycle of guessing and testing, the
blackboard of theory and the workbench of experiment can no longer
encompass all of physics: there is a gap between them.  In this gap,
something else comes into play, in order for symbol-handling agents (people
or perhaps other organisms) to link evidence to particular explanations.
How, then, are we to think about a physical world now recognized as
inseparable from symbol-handling agents in interaction with other parts of
this world?

Based on the recognition that we ourselves as physicists act as
symbol-handling agents, and that we are organisms alive in the physical
world, we wonder what other aspects of behavior, especially biological
behavior, might fruitfully be explained by invoking symbol-handling agents
as elements of description.  {\it Symbols} and {\it symbol-handling agents}
are terms of description available at widely varying levels of detail.  I
may see myself as handling symbols, and I may inquire into evidence of
symbol handling on the part of mitochondria within my cells.  We propose
that symbol-handling agents enter physical explanations.

Mainstream physics, emphasizing particles and fields, has no explicit
vocabulary for symbol handlers or their symbols.  Recently, however, a crack
opened in physics for discussions of agents and symbols, for example in the
work of Fuchs and Schack \cite{caves,fuchs16} and Briegel \cite{briegel1104}.
In these and other current examples, {\it agency} and {\it agent} name a
variety of notions with a complex history, partly in opposition to obsolete
notions of objectivity that traces back at least to Descartes \cite{riskin}.
Some biologically-oriented notions of agents are introduced in biosemiotics
\cite{sharov} and in code biology \cite{code,barbieri2014,barbieri2016}.

What capabilities are to be ascribed to a `symbol handling agent'?  We
think of such an agent as taking steps, one after another, and as equipped
with a memory.  Each ``next step'' of an agent is influenced both by the
contents of its memory and by an inflow of symbols from an environment that
includes other agents, and also by a logically undetermined ``oracle''
external to the agent \cite{or}.  Guesses come from an agent interacting
with an oracle, the workings of which we refrain from trying to
penetrate. How to elaborate this proposal remains open; presumably
investigators will conceive of a variety of expressions of `symbol-handlers
for applications varying from descriptions of viruses to descriptions of
human mentality.

In order to distinguish between what can be computed and what must come
from beyond computation (as guesses from interaction with an oracle), we
imagine the extreme case of an symbol handler that, while open to guessing,
possesses maximal computational capacity. Thus we are unconcerned with
practical limits on computing imposed by limits on memory or by limits on
the rate at which an symbol handler computes, leading us to assume that the
symbol handler has the ultimate computational capability of a Turing
machine.  The Turing machine, however, requires modification to offer a
place for guesses from interaction with an oracle and for communication
with other such machines.

Fortunately for our purposes, in a side remark to his 1936 paper, Turing
briefly introduced an alternative machine called a {\em choice machine},
contrasted with the usual Turing machine that Turing called an a-machine:
\begin{quote}
  If at each stage the motion of a machine \ldots is completely determined by
  the [memory] configuration, we shall call the machine an ``automatic
  machine'' (or a-machine). For some purposes we might use machines (choice
  machines or c-machines) whose motion is only partially determined by the
  configuration \dots. When such a machine reaches one of these ambiguous
  configurations, it cannot go on until some arbitrary choice has been made
  by an external operator.  This would be the case if we were using machines
  to deal with axiomatic systems. \cite{turing}.
\end{quote}
We picture a symbol handler equipped with a c-machine modified to take
part in a communications network by transmitting symbols to other such
machines.  We call the modified c-machine a Choice Machine.  We posit that on
occasion an ``oracle'' writes a symbol onto the scanned square of the Turing
tape of the symbol-handler's Choice Machine {\em privately}, in the sense that the
symbol remains unknown to other symbol handlers unless and until the symbol-handling
agent that receives the chosen symbol reports it to others \cite{aop16T}.
There is no limit to the number of symbols that the oracle can write, and we make no rule of separating data from program, so that what comes from the oracle can influence both data and programs.

\section{Appreciating uncountability}\label{sec:5}
While we previously linked agency to multiple explanations \cite{aop16}, now
we have the proof that the set of explanations of given evidence is {\it
  uncountably} infinite.  The indefinitely extendable symbol-holding tape of
the Choice Machine gives an interesting way to appreciate this uncountable
set, as follows.  For a finite set, one can write down numerical names for
the elements, one after another: element 1, element 2, etc.  For countably
infinite sets one cannot quite do this, but one imagines a correspondence of
names of the elements of the set with the natural numbers.  One thinks ``as
if'' names of the elements could all be written one after another on a Turing
tape, without ever running out of symbol-holding squares of the tape.
A Turing machine, programmed to get to a the name of a particular element,
will always reach it.

By definition, an {\it uncountable set} cannot have the names of its elements
written on even an unlimited tape, so that even in the theoretical limit of
unlimited memory and unlimited computational speed, an agent's Choice Machine
cannot be relied on to eventually arrive at every explanation that exhibits
some property of interest.  To arrive an explanation or even a subset of
explanations requires a reach beyond what can be computed by any
variety of Turing machine. Because the set of potential explanations is
uncountable, when you guess an explanation, you do more than pick from a
list: something is \vspace*{6pt} created. \\
\noindent\vspace*{-6pt}
\textbf{Remarks}:
\begin{enumerate}
\item G\"odel's first incompleteness theorem implies that arithmetic generates an
  infinite list of questions, each answerable by an agent putting forth an
  answer that inserts one of two axioms, extending arithmetic in an unending
  structure of binary choices.  Thus by the diagonal argument, the set of
  extended arithmetics is uncountable.  G\"odel used a fixed countable system
  of symbols to make his proof from which it follows that the set of extended
  arithmetical systems is uncountable.  Similarly, we use a fixed countable
  system to prove that the set of explanations of given evidence is
  uncountable.
 \item By definition, an axiom is logically undetermined and is thus a species
   of a guess made by some agent.  Recalling the metaphor of the `tree of
   assumptions', the axioms that settle G\"odel's logically undecidable
   questions represent uncountable branches (perhaps better said, branchlets)
   of that tree.
\end{enumerate}

\section{Is agency pervasive?}\label{sec:6}
A major major fork in the tree of assumptions can now be seen.  Physicists
exhibit agency, and if physicists are part of the material world then some
physical material is under the influence of agency.  We now ask: is all
physical matter influenced by agency?  Are the stars made out of only dead
matter, which is to say, are they uninfluenced by agency, or not?  If the
stars are shaped partly by agency, that is, by acts of agents, agents that
are by definition capable of uncomputable behavior, a new area for
investigation opens up.  Habit in physics weighs against this view, but that
habit depends on the notion that although ``now we see through a glass
darkly,'' there is some ultimate end picture toward which we approach.  In
contrast, the implication of the proof of uncountable explanations of any
given evidence, along with the recognition of symbol-handling as part of the
material world, decides for the other major branch: a physical world pervaded
by agency.

In a world pervaded by agency, the big-bang theory still serves to explain a
body of cosmological evidence; we do not criticize that explanation, but we
recognize that there is potentially an uncountable set of other explanations
that also explain the evidence.  One no longer can hope to relate the world
of physical experience to a single system of logic, comparable to what
G\"odel used to prove the incompleteness theorems, but rather the world of
physical experience is to be seen as something for which explanations, like
extended arithmetics, form an uncountable  set.

\section{Discussion}\label{sec:7}
Proven above is the Theorem that all explanations are necessarily open to
testing and to needs for revision, so that any path of inquiry remains
uncloseably open; there can be no final answers compatible with quantum
mechanics.  How is one to respond?  One is left with a question of faith. Is
the conflict of quantum theory with final answers something to respond to
with despair or with joy?  That choice no science of which we know can
resolve: it is up to the person.

For a next step along our path of inquiry, we plan to follow up on the
expression of evidence and its explanations by flows of strings of symbols in
cycles of guessing and testing. It might be supposed symbols flow within some
fixed spacetime.  On the contrary, however, we are interested in the
establishment, maintenance, and dissolution of flows symbols among agents as
enabling an uncountable set of other systems of spatio-temporal management,
differing from the spacetimes of special or general relativity, systems of
spatio-temporal management that accommodate to the circumstances in which the
agents operate.  For example, animal behavior, as in E. Coli
\cite{selfridge}, frog vision \cite{lettvin}, and depth vision of praying
mantis \cite{mantis} come to mind as places to start to look for interesting
systems of spatio-temporal \vspace*{8pt} management.

Readers of an earlier draft offered several comments and criticisms deserving
of the following responses.

\subsection{Non-triviality of tensor products}
Is a density operator that is a tensor product trivial? One can always write a
density operator for each of two unrelated experiments as a tensor product
with one factor for each experiment, and it is fair to call such a tensor
product `trivial'; however, products of density operators enter explanations
non-trivially, for example when two parameters of a single experiment enter
distinct factors of a tensor product of density operators. In that case a
choice of a parameter can have an effect that is unexpressed in a simpler
explanation, but has a dramatic effect when a tensor product is invoked in a
more encompassing explanation.

For example, this situation occurs in the BB84 protocol for Quantum Key
Distribution (QKD), which relies on quantum uncertainty to claim security
against undetected eavesdropping.  In BB84\cite{BB84}, Alice transmits a
sequence of laser-generated, single-photon light pulses to Bob, from which
Bob and Alice extract a key.  For each pulse, Alice ``sets a knob'' to choose
randomly among four types of light pulses, each characterized by a
polarization vector and a frequency spectrum.  If the spectra are identical,
the experiment can be explained by a density operator on the two-dimensional
vector space of polarizations, as assumed in much of the QKD literature.  In
a two dimensional space, there must be substantial overlap among the four
polarizations, forcing substantial uncertainty on any eavesdropper's
measurements.  The claim of QKD security depends on this uncertainty.
However, a physically more informed explanation accounts for light frequency,
for which it introduces another factor, of infinite dimension, so that the
density operator becomes a tensor product of a density operator expressing
polarization with a density operator expressing the frequency spectrum of the
light pulse.  By ``setting a knob'', Alice determines both a polarization
angle {\it and a frequency}.

The simpler explanation, employing the density operator on the
2-dimensional polarization space, gives an adequate description of
eavesdropping under the condition that the eavesdropper makes no use of
frequency discrimination. But the tensor-product explanation is critical to
showing what a better equipped eavesdropper can learn.  In some popular
implementations, each type of light pulse comes from a different laser.  It
is readily shown that if the lasers are imperfectly aligned in frequency, the
overlaps among the four light states approach zero, so that the QKD system is
seen have no security at\vspace*{6pt} all\cite{10CUP}.

\subsection{Entanglement}
Here is an example of two explanations of given evidence, the first of which
involves no entanglement, while the second involves an an entangled state.
Consider a simple case of evidence with two outcomes: $\omega^{(0)}_1$ and $\omega^{(0)}_2$, and with probabilities given, for some $0\le a \le 1$, by
\begin{equation}\label{eq:prob}
  \Pr(\omega^{(0)}_1)= a,\;\text{and}\; \Pr(\omega^{(0)}_2)= 1-a.
\end{equation}
These probabilities accord with an explanation $\alpha$
involving a Hilbert space of just two real dimensions, $\mathcal{H}^{(0)}=
R^2$, with a basis $\{|x^{(0)} \rangle,|y^{(0)}\rangle\}$, so that
\begin{equation}
  |x^{(0)}\rangle\langle x^{(0)}|+|y^{(0)}\rangle\langle y^{(0)}| = \bm{1}^{(0)},
\end{equation}
with
\begin{eqnarray}
  \rho_\alpha &=& (\sqrt{a}|x^{(0)}\rangle+\sqrt{1-a}|y^{(0)}\rangle)(\sqrt{a}\langle x^{(0)}|+\sqrt{1-a}\langle y^{(0)}|) ;\\
  M_\alpha^{(0)}(\omega^{(0)}_1)&=&|x^{(0)}\rangle\langle x^{(0)}| \;\text{and}\;\\
  M_\alpha^{(0)}(\omega^{(0)}_2)&=&|y^{(0)}\rangle\langle y^{(0)}|
\end{eqnarray}
An alternate explanation $\beta$ of the probabilities given in
(\ref{eq:prob}) asserts an entangled state.  The explanation $\beta$ involves
an additional tensor-product space $\mathcal{H}^{(1)}=R^2$, again a real
vector space of dimension 2, with basis vectors $\{|x^{(1)}\rangle,
|y^{(1)}\rangle\}$:
\begin{eqnarray}
  \rho_\beta &= & \left(\sqrt{a}|x^{(0)}\rangle|x^{(1)}\rangle + \sqrt{1-a}|y^{(0)}\rangle |y^{(1)}\rangle\right) \nonumber\\ && \;\: \left(\sqrt{a}|\langle x^{(0)}|\langle x^{(1)}|+\sqrt{1-a}\langle y^{(0)}|\langle y^{(1)}|\right)\\
  M_\beta&=&M_\alpha^{(0)}\otimes \bm{1}^{(1)}
\end{eqnarray}

\subsection{Concept of an explanation involving an infinite string of symbols}
Can it make sense to admit explanations expressed as (countably) infinite
strings of symbols? One can view (\ref{eq:two}) as a mapping from infinite
binary sequences to explanations.  Looked at that way, an explanation is then
a particular infinite bit sequence together with the mapping.  (The mapping
is expressed by a finite number of symbols (it fits on the page!), but the
bit sequence is (countably) infinite.)  What leads us to propose such a form
as an explanation is the conceptual separation between theory on the
blackboard and experimental activity on the workbench: we no longer see
practical constraints on the workbench as constraining the forms admissible
for theory.  A precedent for the lack of constraint is the Turing machine
with its infinite tape.
  
The separation between explanations on the blackboard and evidence on the
workbench established by the Theorem is already provided by a weaker theorem
in \cite{aop05} which implies a countable (not an uncountable) set of
explanations of given evidence.  This multiplicity shows that wave functions
and linear operators cannot be found on the workbench of evidence, but
instead are imaginative entities to be seen on the blackboard of mathematics.
What to make of this separation of explanations from anything that can be
seen on the workbench?  From a standpoint that recognize to agency,
explanations on the blackboard appear as strings of symbols written by
agents, and are acceptable entities toward which to devote theoretical
attention.  So, if one wants to build a mathematical theory of explanations,
what should be the building blocks?

To choose the ``building blocks'' for such a theory, it may help to notice
the use of number systems in physics.  In that use one tolerates a separation
between the mathematical properties by which number systems are defined and
experience with physical acts of computation.  The limited memory of any
actual computer puts a limit on the range of integers that the computer can
deal with one-to-one, so the computer cannot represent the whole set of
integers.  Instead it represents a subset of integers, and appends to this
subset some way of dealing with underflow and overflow.  Besides the
well-known issue of round-off errors in floating-point representations of the
integers, one gets a second effect, more interesting for present purposes.
For arithmetic operations on some integers $x$, $(x+x)-x$ overflows, while
$x+(x-x)$ does not overflow, but gives $x$; the associative law of addition
suffers exceptions.  (Try it in Matlab for $x= 10^{308}$).  When dealing with
integers near the limits of the capacity of the computer, a programmer has to
deal with this and other exceptions to the ``laws of arithmetic.''  Should
physicists use those laws or should they instead use the much more
complicated behavior of expressions of number as actually implementable in
computer hardware on the workbench?  For physics, the answer is driven by the
drive for elegance and simplicity: ``An important task of the theoretical
physicist lies in distinguishing between trivial and nontrivial discrepancies
between theory and experiment.''\cite[ p3]{feshbach} The mathematical idea of
integers---that every integer has a successor and a predecessor---so charms
the mind that the gap between the mathematically defined integers and their
limited realization in computers is no fatal stroke against their use in the
constructing of theory.

Now back to building blocks for a theory of explanations.  We start with a theoretical abstraction, parallel to the that of the integers: in explaining given evidence, one can always take one more factor into account.  Do we accept this abstraction or not?   Were we to hew too closely to what can be realized, we expect to eventually bog down if we keep trying to add factors realizable on the workbench; however, to account for this limitation of the workbench, we must deal with the  vexing question of ``what is the maximum number of factors?'', which depends on the technology on hand and, and so evolves with that technology.  For theoretical purposes we ignore such a limit, which, once done, opens the way to theoretical portrayals of explanations as in (\ref{eq:two}).

\subsection{Testing extended explanations }
A question was raised about testing extensions of explanations of the form
(\ref{eq:two}).  Before they are extended, those explanations (\ref{eq:two})
are no harder to test than is the explanation (\ref{eq:one}), for the simple
reason that all these explanations, with the extra factors of the POVM set to
unit operators ($M^{(n)}(\Omega^{(n)})=\bm{1}^{(n)}$), assert exactly the
same probabilities.  The extended explanations obtained by attending to finer
outcomes $\omega^{(n)}\subset\Omega^{(n)}$ for some values of $n$ are
testable in principle, to the extent of selecting any particular factor, say
the $m$-th.  To test the extended explanation with respect to the $m$-th
factor, all the factors $n\ne m$ are ignored, as expressed by unit POVM
factors, so that (\ref{eq:two}) reduces to
\begin{equation} \label{eq:red}
  \left(\rho^{(0)}\otimes \rho^{(m)}_{j_m}, M^{(0)}(\omega^{(0)})
  \otimes M^{(m)} (\omega^{(m)})\right),
\end{equation}
The trace of (\ref{eq:red}) is the marginal probability obtained by ignoring
outcomes of all sectors except sector 0 and sector $m$, and the distinctions
in probability do not become small as $m$ increases.

\subsection{Restriction to finite sets}
Some readers may still be uncomfortable with explanations involving infinite
tensor products, so we address what remains of our argument if only finite
sets are admitted.  Suppose that we allow only some finite number $N$ of
tensor-product factors, and, correspondingly, only $N$ extensions of
arithmetic.  Also, we truncate the Turing tape to make a computer of finite
memory capacity.  Then there are $2^N$ extended arithmetics.  Similarly, the
number of alternative explanations is $\ge 2^N$.  Thus, recording the names
of all the alternative explanations or of all the extended arithmetics places
a demand on computer memory that increases exponentially with $N$.  The
parallel between G\"odel incompleteness and explanations is still present in
the exponential behaviors, but, with the restriction to finite sets, the
parallel is expressed by {\it complexity}, rather than by the (to us)
mathematically more appealing formulation in terms of computability.

\section*{Acknowledgment}
We thank Kenneth Augustyn for helpful comments.
We thank Louis Kauffman, Paul Benioff, and an anonymous critic for
pointing us to issues discussed in Sec.\ \ref{sec:7}.

\end{document}